\documentclass{article}
\usepackage{xcolor}

\usepackage{microtype}
\usepackage{graphicx}
\usepackage{booktabs}
\usepackage{amsmath,amssymb}
\usepackage{xurl}
\usepackage{hyperref}
\usepackage[accepted]{mlsys2025}
\usepackage{etoolbox}
\makeatletter
\renewcommand{\Notice@String}{Preprint.}
\makeatother
\patchcmd{\printAffiliationsAndNotice}
  {Correspondence to:}{}{}
  {\PackageError{author-block}{Could not update email label}{}}
\hypersetup{pdfsubject={arXiv preprint}}

\usepackage{framed}
\definecolor{formalshade}{rgb}{0.95,0.95,0.97}
\definecolor{darkblue}{rgb}{0.14,0.22,0.52}

\newcounter{takeaway}
\renewcommand{\thetakeaway}{\arabic{takeaway}}

\newenvironment{takeaway}{
  \refstepcounter{takeaway}
  \def\FrameCommand{
    \hspace{-4.55pt}
    {\color{darkblue}\vrule width 1.5pt}
    \colorbox{formalshade}
  }
  \MakeFramed{\advance\hsize-\width\FrameRestore}
  \noindent\hspace{-4.55pt}
  \textbf{Takeaway~\thetakeaway.}\enspace
}
{
  \endMakeFramed%
  \vspace{-2pt}
}

\usepackage{enumitem}

\renewcommand{\algorithmicrequire}{\textbf{Input:}}
\renewcommand{\algorithmicensure}{\textbf{Output:}}

\newcommand{\sysenergy}{\mathcal{E}_{\mathrm{GPU}}}

\begin{document}
\twocolumn[
% \mlsystitle{Energy--Performance Trade-offs in Multimodal LLM Serving:\texorpdfstring{\\}{ } EPD Disaggregation and Stage-Wise DVFS}
\mlsystitle{A Matter of Stages: Energy--Latency Trade-offs in Multimodal LLM Serving}
\begin{mlsysauthorlist}
\mlsysauthor{Yifan Sun}{unimelb}
\mlsysauthor{Jiantong Jiang}{unimelb}
\mlsysauthor{Zhexiang Zhang}{unimelb}
\mlsysauthor{Xu Bai}{unimelb}
\mlsysauthor{Feng Liu}{unimelb}
\mlsysauthor{Rajkumar Buyya}{unimelb}
\mlsysauthor{Adel N. Toosi}{unimelb}
\end{mlsysauthorlist}
\mlsysaffiliation{unimelb}{University of Melbourne, Melbourne, Australia}
\mlsyscorrespondingauthor{Yifan Sun}{\nolinkurl{yifan.sun.4@student.unimelb.edu.au}}
\mlsyscorrespondingauthor{Jiantong Jiang}{\nolinkurl{jiantong.jiang@unimelb.edu.au}}
\mlsyscorrespondingauthor{Zhexiang Zhang}{\nolinkurl{zhexiang.zhang@student.unimelb.edu.au}}
\mlsyscorrespondingauthor{Xu Bai}{\nolinkurl{xu.bai1@student.unimelb.edu.au}}
\mlsyscorrespondingauthor{Feng Liu}{\nolinkurl{feng.liu1@unimelb.edu.au}}
\mlsyscorrespondingauthor{Rajkumar Buyya}{\nolinkurl{rbuyya@unimelb.edu.au}}
\mlsyscorrespondingauthor{Adel N. Toosi}{\nolinkurl{adel.toosi@unimelb.edu.au}}
\mlsyskeywords{Multimodal inference, energy measurement, disaggregated serving, GPU frequency scaling}
\vskip 0.3in
\begin{abstract}
Online serving of multimodal large language models requires deciding whether to separate visual encoding (E), prefill (P), and decode (D) across GPUs and how to configure their frequencies. EPD disaggregation enables independent stage control, but it also changes communication and queueing behavior. Lowering one stage’s frequency can increase queueing and prolong execution across the EPD pipeline. Consequently, reducing power at one stage does not necessarily reduce total GPU energy per request. In this paper, we systematically characterize fused and EPD-disaggregated serving across four multimodal models under a fixed three-GPU budget, followed by single-stage and uniform downclocking experiments across 18 workload combinations. We find that disaggregation reduces GPU energy per successful request by 13.3–37.0\% at nearly identical output-token throughput, but increases mean TTFT. Downclocking exhibits distinct stage sensitivities, and aggressive frequency reductions can offset power savings through queueing and longer execution. Across the evaluated workloads, uniform downclocking to 80\% reduces GPU energy by 11.7–18.6\% with modest mean-latency changes. Guided by these findings, we develop a budgeted frequency exploration strategy to select energy-efficient configurations under latency constraints. Evaluation shows that four probes can identify the lowest-energy feasible configuration among 25 measured candidates across all 288 evaluated positive-margin SLO scenarios. These results provide practical guidance for energy-efficient multimodal serving.
\end{abstract}
]
% The template notice is an unnumbered footnote without a hyperlink target.
\begin{NoHyper}
\printAffiliationsAndNotice{\urlstyle{same}\Urlmuskip=0mu\relax\def\UrlBreaks{\do\.\do\@}\def\UrlBreakPenalty{0}\hyphenpenalty=10000\relax}
\end{NoHyper}

\section{Introduction}
\label{sec:intro}

\begin{figure*}[t]
  \centering
  \includegraphics[width=\textwidth]{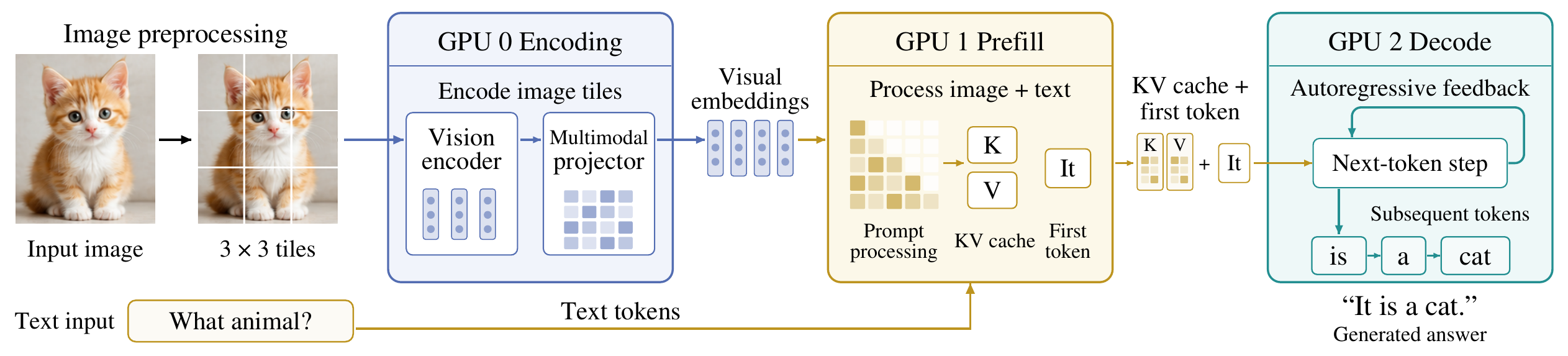}\par\nointerlineskip
  \caption{EPD-disaggregated inference for an image--text request. Encoding produces visual embeddings. Prefill builds the KV cache and generates the first token. Decode generates subsequent tokens autoregressively.}
  \label{fig:epd-pipeline}
\end{figure*}

Multimodal large language models combine visual understanding with language generation, supporting applications such as visual question answering, document understanding, and chart analysis~\cite{idefics2,docowl2,chartgemma}. Serving these models online must balance response latency with GPU energy consumption. 
%meet users' responsiveness requirements. 
An image-text request typically passes through visual encoding (E), language-model prefill (P), and autoregressive decode (D), as illustrated in Fig.~\ref{fig:epd-pipeline}. These stages differ in computational demand, execution time, and resource use, making their organization and provisioning important to serving performance.
%How they are organized and provisioned directly affects serving performance. Meanwhile, the GPUs supporting online inference incur substantial energy costs. 
Recent inference-energy research further shows that GPU energy required to process a workload depends on serving configurations~\citep{mlenergy}, motivating evaluations that consider energy alongside latency. 
%serving configurations affect the energy required to complete the same inference workload, motivating deployment decisions that consider energy alongside performance~\cite{mlenergy}.

Disaggregated serving provides a way to organize resources around the different demands of inference stages. DistServe~\cite{distserve} and Splitwise~\cite{splitwise} place language-model prefill and decode to reduce interference and enable independent resource provisioning. For multimodal inference, 
%Multimodal serving introduces visual encoding as an additional computational stage. 
EPD disaggregation further separates visual encoding from language-model computation, forming a serving pipeline with independent E, P, and D instances~\citep{epd_icml, epdserve_ascend}.

GPU frequency is another configuration choice that affects energy and performance. Dynamic voltage and frequency scaling (DVFS) has been used to optimize inference energy by adjusting processing capacity and reducing power within latency requirements. DynamoLLM~\cite{dynamollm} and BEAM~\cite{beam} study the joint selection of frequencies and other serving configurations, demonstrating the potential to use performance slack for energy savings. In multimodal EPD serving, separating stages allows E, P, and D to operate at different frequencies, but also changes how a fixed GPU budget is allocated and introduces cross-GPU transfers and separate queues. Thus, their  %EPD disaggregation enables independent stage execution and different GPU frequencies for different stages. However, it also introduces state transfers, stage handoffs, and new queueing behavior. Under a fixed GPU budget, stage isolation and resource concentration jointly affect serving performance, so 
energy benefits must be assessed via measurements of the complete pipeline.

However, the energy–performance trade-offs of a complete multi-GPU EPD pipeline remain insufficiently characterized.
%Research on energy consumption in multimodal EPD serving is still at an early stage, and the energy--performance trade-offs of the full multi-GPU pipeline have yet to be fully characterized. 
\citet{sc25_stage_energy,modality_inflation} profile stage-level energy consumption in single-GPU multimodal inference, while \citet{revisiting_energy} investigate energy and DVFS in PD disaggregated LLM serving. However, these studies do not characterize the combined effects of E, P, D, and their interactions in a multi-GPU EPD pipeline. Stage separation changes GPU utilization, cross-GPU transfers, and queueing behavior. 
%separately profile visual encoding and prefill on a single GPU, examining how model architecture and input complexity affect energy consumption and exploring stage-specific DVFS.
%inference latency. These measurements inform stage-specific DVFS and input-aware batching. 
% However, this single-GPU setup does not capture interactions among E, P, and D running concurrently on separate GPUs. 
Downclocking one stage may prolong execution of other stages in the pipeline, and  
%request processing and benchmark execution, causing other GPUs to consume energy for longer. 
savings in a stage with a small energy share may have limited overall impact. Thus, reducing stage power 
%Savings at one stage may also have limited overall impact if that stage accounts for only a small fraction of total energy. A reduction in stage power therefore 
does not directly establish a reduction in energy per request for the complete service.

% To clarify these whole-pipeline energy--performance trade-offs and inform deployment and stage-frequency selection, this paper addresses two related research questions.
These limitations motivate two key research questions:

\textbf{RQ1.} \textit{When does EPD disaggregation reduce GPU energy compared with fused serving under the same GPU budget?}

% \noindent\textit{(1) When should multimodal online serving use fused deployment or EPD disaggregation?}

\textbf{RQ2.} \textit{Once EPD is disaggregated, how should stage frequencies be adjusted to reduce whole-pipeline GPU energy under latency constraints?}

To address these questions, we compare three fused replicas with a 1E+1P+1D pipeline and then evaluate single-stage and uniform downclocking in the disaggregated service. We further use the resulting stage-level findings to guide frequency exploration under a user-specified profiling budget. Our main contributions are as follows.

\begin{itemize}[leftmargin=*,itemsep=2pt,topsep=3pt,parsep=0pt]
  \item We characterize the energy--latency trade-offs between fused and disaggregated serving across four models under the same three-GPU budget. Disaggregation reduces GPU energy per successful request by 13.3--37.0\% at nearly identical output-token throughput, but increases mean time to first token (TTFT). Its effect on time per output token (TPOT) depends on the model and request rate.
  \item Through extensive experiments across 18 workload combinations, we identify distinct stage responses to downclocking. E offers limited whole-pipeline savings, deep P downclocking can trigger queueing and prolong execution, and moderate D downclocking provides useful savings with small TPOT penalties. Moderate uniform downclocking is also effective. An 80\% frequency target reduces total GPU energy by 11.7--18.6\% relative to the default-frequency disaggregated baseline, with small changes in mean latency. The performance boundaries vary with model, request rate, and workload content.
  \item We develop a budgeted frequency exploration algorithm that uses these findings and measured energy and latency feedback to guide configuration selection. Our evaluation shows that four probes identify the lowest-energy feasible configuration among 25 measured candidates across all 288 evaluated positive-margin SLO scenarios.
\end{itemize}

\section{Background and Motivation}
\label{sec:background}

\subsection{Multimodal Large Language Models}
\label{sec:bg-models}
% Modern large language models increasingly support images alongside text, enabling visual question answering, image description, and image-grounded dialogue. These multimodal models typically connect a vision encoder to a language model through a learned adapter or projection~\cite{visualinstruction}. 
Multimodal large language models (MLLMs) extend language models with vision encoders to process images alongside text~\citep{visualinstruction, blip2}. In a common architecture, an image is preprocessed into patches, potentially spanning multiple image tiles, and transformed into visual features by a vision encoder. These visual features are projected or integrated into the language model for image-grounded text generation. Figure~\ref{fig:epd-pipeline} illustrates this process using a cat image and the question ``What animal?''
%. The vision encoder transforms these patches into visual features, which are mapped into representations that the language model can process together with the text prompt. Through multimodal training, the model learns to relate these visual representations to language and generate responses grounded in the image.

For the image--text-to-text models considered in this work, inference comprises \emph{encoding} (E), \emph{prefill} (P), and \emph{decode} (D). E extracts visual representations from input images. P processes the visual and textual inputs, builds the key--value (KV) cache, and generates the first output token. D then reuses and extends the cache to generate subsequent tokens autoregressively~\cite{epd_icml}. %Image resolution, image count, and model-specific tokenization determine the amount of visual processing and the context passed to the language model.

The computational demand of these stages depends on both the model architecture and the input workload. Image resolution affects encoding cost, while model-specific visual-token processing influences prefill computation and KV cache requirements. Decode workload further depends on context and output length. Consequently, MLLMs with similar parameter counts can exhibit different stage-level serving demands~\citep{sc25_stage_energy,modality_inflation}.

\subsection{EPD-Disaggregated Serving}
\label{sec:bg-disaggregation}
Prefill-decode disaggregation has emerged as a general approach to improving language-model serving. %Systems such as DistServe~\cite{distserve} and Splitwise~\cite{splitwise} 
It places prompt processing and token generation on separate GPU resources, allowing the two phases to be provisioned and scheduled independently~\cite{splitwise,distserve,mooncake}. This separation reduces interference between prefill and ongoing decoding, whose execution patterns and latency requirements differ.

Multimodal serving introduces visual encoding as an additional stage, motivating EPD disaggregation~\citep{epd_icml, triinfer}.
%As multimodal inference introduces visual encoding before prefill, this approach extends naturally to separating the encoder. EPD serving systems isolate encoding from language processing and support independent stage scheduling and resource allocation~\cite{epd_icml,triinfer}. 
In a fully separated EPD pipeline, E sends visual representations to P, and P transfers the resulting KV cache to D. The three instances can process different requests concurrently, although each request must traverse the stages in order. The GPUs can also be configured independently. Figure~\ref{fig:epd-pipeline} illustrates this inference flow.
However, disaggregation introduces communication and coordination overheads and changes GPU resource contention. These effects influence \emph{time to first token} (TTFT) and \emph{time per output token} (TPOT), making the benefits of disaggregation dependent on the model, workload, and resource allocation.
% End-to-end performance is reflected in time to first token (TTFT), which measures initial responsiveness, and time per output token (TPOT), which measures the pace of subsequent generation. Separation introduces communication and coordination overheads, so its performance and energy benefits depend on the workload and resource allocation.

\subsection{GPU Energy and DVFS}
\label{sec:bg-gpu-energy}
Online LLM serving relies on GPUs whose sustained operation incurs substantial energy costs~\cite{mlenergy}. Reducing GPU energy while maintaining service capacity and meeting request latency requirements is therefore increasingly important~\cite{dynamollm}. Besides matching active GPU resources to demand, GPU frequency scaling provides another means of power management~\cite{slo_gpu_dvfs}. Dynamic voltage and frequency scaling (DVFS) adjusts processor frequency and associated operating voltage to balance performance and power. Lower frequencies can reduce power but may prolong execution. Energy savings depend on whether the power reduction compensates for the longer runtime~\cite{beam}. When latency budgets provide sufficient slack, systems can use that slack to downclock GPUs and save energy while meeting the application's service requirements.

\subsection{Motivation}
\label{sec:bg-gap}

Prior work has extensively studied disaggregated serving to improve LLM inference efficiency~\citep{kumar2025attention,skis}.
%Prefill--decode (PD) disaggregation~\cite{distserve,splitwise,mooncake} has become an important architectural choice for online LLM serving. By placing prefill and decode on separate resources, it allows parallelism strategies, hardware, and resource allocation to be tailored to their different compute, memory-access, and latency requirements, while reducing the interference associated with co-scheduling chunked prefill and ongoing decoding on the same GPU. Meanwhile, as generative AI services scale, the GPU clusters supporting inference incur substantial energy costs~\cite{mlenergy}, making energy consumption and power budgets important constraints on online service deployment~\cite{dynamollm,beam}.
EPD-Serve~\citep{epdserve_ascend} and TriInfer~\citep{triinfer} optimize multimodal EPD serving through stage-specific resource allocation and scheduling. \citet{mitra2026beyond} characterize PD disaggregation and show that its benefits depend on workload characteristics and resource allocation. These studies establish the performance opportunities of disaggregation but do not examine its energy–latency trade-offs under stage-specific frequency scaling in multimodal EPD serving.

\begin{figure}[t]
  \centering
  \includegraphics[width=\columnwidth]{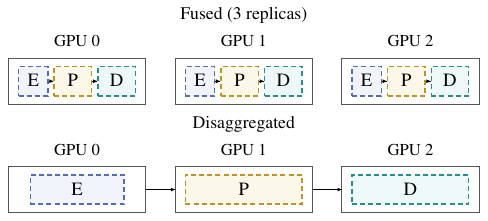}%\par\nointerlineskip
  \caption{Fused and EPD-disaggregated execution.}
  \label{fig:fused-epd-deployment}
\end{figure}

Energy-oriented studies provide complementary insights. \citet{revisiting_energy} investigate the performance and energy implications of PD disaggregation and GPU frequency scaling in LLM serving, finding that their evaluated disaggregated configurations consume more energy than colocated serving. In multimodal EPD serving, 
%Energy characterization of multimodal EPD serving is still at an early stage. The study by 
\citet{sc25_stage_energy,modality_inflation} examine stage-level energy costs using a single NVIDIA A100. These measurements help explain where energy is spent within a request, but the analysis remains limited to inference stages on the same GPU. %The authors also identify multi-GPU disaggregation as a direction for future work. 
Figure~\ref{fig:fused-epd-deployment} illustrates the architectural distinction. In fused execution, a request passes through E, P, and D on the same GPU. In disaggregated execution, the three stages run on separate GPUs as independent serving instances that can process different requests concurrently. Therefore, the single-GPU stage measurements fail to capture the energy behavior of the complete disaggregated EPD pipeline, including cross-GPU transfers, queueing, and interactions between stages.

These cross-stage interactions also affect the energy impact of DVFS. Downclocking one stage can alter queueing and delay downstream execution, prolonging the period during which other GPUs continue drawing power. Consequently, lower power at one stage does not necessarily translate into lower total GPU energy.
%The energy behavior of such a deployment cannot be inferred directly from isolated stage measurements. Transfers and queues change when each GPU computes or waits, and reducing one stage's frequency can affect the execution of downstream stages. A lower power draw at one GPU may therefore be offset by a longer workload completion time and continued energy consumption elsewhere in the pipeline. 
For a three-GPU EPD-disaggregated deployment, total GPU energy is
\begin{equation}
\sysenergy=\int_{t_0}^{t_1}\!\left(P_E(t)+P_P(t)+P_D(t)\right)\,dt,
\label{eq:bg-pipeline-energy}
\end{equation}
%where the common measurement interval includes both computation and waiting.
where $P_E(t)$, $P_P(t)$, and $P_D(t)$ denote the instantaneous power of the GPUs assigned to encoding, prefill, and decode, respectively. The common measurement interval $[t_0,t_1]$ includes both computation and waiting on all three GPUs.
Evaluating a frequency change consequently requires measuring its effect on the complete pipeline, rather than only the power or energy of the modified stage.

These limitations motivate two complementary questions.

\textbf{RQ1.} \textit{Considering both energy and performance, when should multimodal serving use EPD disaggregation, and when should it use fused deployment?}

We examine whether separating E, P, and D decreases or increases total GPU energy relative to fused replicas, and whether separation adds overheads or improves serving performance. Answering RQ1 requires comparing both deployments under the same GPU budget and request workload, and examining how the energy--latency trade-off changes across models and request rates.

\textbf{RQ2.} \textit{Once encoding, prefill, and decode are disaggregated, how should stage-wise DVFS be applied to reduce whole-pipeline energy under latency constraints?}

% This requires characterizing how individual and joint frequency reductions affect total GPU energy, TTFT, and TPOT while all stages operate together. Request rate and workload content may change which stage becomes a bottleneck and how much downclocking is tolerable. Our study therefore examines both deployment selection and frequency selection, aiming to identify where energy can be saved, what performance costs accompany those savings, and when further downclocking ceases to be beneficial.

To investigate stage-wise DVFS (RQ2), we examine three subquestions. RQ2.1 examines how single-stage and uniform downclocking affect whole-pipeline GPU energy and end-to-end performance, measured by TTFT and TPOT. RQ2.2 investigates how request rate and workload content change these energy--performance trade-offs and the scope for downclocking under latency constraints. RQ2.3 examines whether the observed stage sensitivities and energy--performance trends generalize across multimodal models. Together, these subquestions establish when and where downclocking is beneficial, how the benefits of stage-wise downclocking depend on the workload, and which findings can guide frequency selection beyond a single model.

\section{Experimental Methodology}
\label{sec:method}
Our experiments address the two research questions introduced above. First, we compare fused and EPD-disaggregated serving under the same three-GPU budget, measuring GPU energy and serving performance. Second, we study stage-wise frequency scaling within the disaggregated pipeline to examine how its energy--performance trade-offs vary with the stage being downclocked, request arrival rate, workload content, and model.

\subsection{Testbed}
\label{sec:method-testbed}
We run all the experiments within a Dell PowerEdge server equipped with four NVIDIA L40S GPUs, each with 48\,GB of device memory, 1\,TB of host memory, and two Intel processors with 32 physical cores each (64 cores in total). We use three GPUs for inference. The fourth GPU is excluded from serving-energy accounting. The GPUs are PCIe-attached. For disaggregated communication, we use the NIXL/UCX stack integrated with vLLM, with reliable-connected RDMA and CUDA-copy transports enabled.

\subsection{Fused and Disaggregated Serving Frameworks}
\label{sec:method-framework}
We use vLLM~\cite{vllm} as the inference engine. The fused deployment consists of three independent replicas, each executing E, P, and D on one GPU. A proxy distributes requests to the replicas in round-robin order.
The disaggregated deployment assigns one GPU to each stage (1E+1P+1D). E supplies visual representations to P, and P transfers KV state to D. Both deployments receive the same request workload at each operating point. All instances use eager execution with prefix caching disabled. Each serving instance uses a single GPU.
We use vLLM~0.30.0 that supports asynchronous GPU--CPU copies, which can reduce the latency of transferring visual tokens to the prefill instance after disaggregation and reduce experimental bias.

For DVFS, we lower E, P, or D individually, or all three stages uniformly. Using the L40S maximum frequency of 2,520\,MHz (100\%) as the reference, we control the frequency at six targets, 90\%, 80\%, 70\%, 60\%, 50\%, and 40\%. Together with the default configuration, these four scaling modes and six frequency levels yield 25 candidate configurations per workload. Although the L40S supports graphics-clock frequencies as low as 210\,MHz~\cite{nvidia_l40s_clocks}, some tests already exhibit request accumulation and inference-framework crashes at the 40\% frequency target. To maintain the validity of the experiments, we therefore set the minimum frequency to 40\% of the maximum. Following $\mu$-Serve~\cite{muserve}, we focus on GPU core frequency scaling. The memory frequency is not adjusted and remains at its default setting throughout each sweep.

\subsection{Models}
\label{sec:method-models}
We select four image--text models to cover contrasting demands on visual encoding, language processing, and KV-cache storage and transfer during online serving.

\textbf{InternVL3-9B}~\cite{internvl} supports dynamic image sizes by dividing images into 448$\times$448 tiles. Its approximately 300M-parameter InternViT encoder processes multiple tiles per image, generating a very large number of visual tokens and imposing a heavy encoder workload.

\textbf{Llama-3.1-Nemotron-Nano-VL-8B-V1}~\cite{nemotron} combines a C-RADIOv2-H vision encoder with Llama-3.1-8B and supports 512$\times$512 image tiles. This is another model with substantial encoding demand.

\textbf{Qwen3-VL-8B-Instruct}~\cite{qwen} combines dynamic-resolution visual processing, spatial token merging, and DeepStack integration of visual features with an 8B language backbone. Its visual-processing demand is substantially lower than that of the preceding two models, resulting in a more balanced workload across stages.

\textbf{LLaVA-1.5-7B}~\cite{llava} combines CLIP ViT-L/14 with Vicuna-7B~\cite{vicuna_config} and produces 576 visual tokens from a 336$\times$336 image. This limits visual processing compared with multi-tile inputs. Its language backbone uses multi-head attention with 32 KV heads per layer, giving a larger per-token KV-cache footprint than other evaluated models at equal cache precision. It contrasts relatively light encoding with greater KV storage and transfer demands.

We use all four models for the deployment comparison in RQ1. Since Nemotron and InternVL3 share a similar workload characteristic, we retain Nemotron, Qwen3-VL, and LLaVA for the stage-wise DVFS experiments in RQ2.

\subsection{Datasets}
\label{sec:method-datasets}
We use two image--text workloads.

\textbf{VisionArena-Chat}~\cite{visionarena} contains real users' image--text conversations with vision--language models.

\textbf{Chart-to-text}~\cite{charttotext} uses the \texttt{chart2text(cauldron)} subset of LLaVA-OneVision-Data~\cite{onevisiondata}, comprising chart images and the text tasks associated with those images.

\subsection{SLOs and Request-Rate Selection}
\label{sec:method-slo}
For the DVFS experiments addressing RQ2, we define latency budgets to guide request-rate selection. Following EPD-Serve~\cite{epdserve_ascend}, we set the P99 TTFT and mean TPOT budgets to 2\,s and 50\,ms, respectively. Based on these budgets, we select an upper request-rate reference for each model--dataset pair and progressively lower the request rate to examine performance and energy trends.

\subsection{Request Generation and Energy Measurement}
\label{sec:method-measurement}
We adjust the number of requests according to the request rate so that each experiment lasts approximately 600--700\,s. Within each DVFS suite, all frequency settings use the same total number of requests for comparison. Requests use exponentially sampled interarrival intervals normalized to the target rate. Within each DVFS workload, frequency settings reuse the same ordered requests and a fixed seed for arrival generation.

We measure the combined board energy of the three serving GPUs using NVML~\cite{nvml}. From the start to the end of the benchmark, we record the power of all three GPUs once per second and integrate them to obtain the total energy consumed. The benchmark continues until all submitted requests finish, and energy measurement includes the final queue drain.
To avoid bias, we schedule three repetitions for each operating point in the frequency-controlled DVFS experiment matrix and report the average.

\subsection{Evaluation Metrics}
\label{sec:method-metrics}
We evaluate performance using output-token throughput, TTFT, and TPOT. Output-token throughput is the number of generated output tokens from successful requests divided by benchmark duration, in tokens/s. TTFT measures the interval from sending a request to receiving its first output token. TPOT measures the average time to generate each subsequent output token after the first.
For energy efficiency, we report GPU energy per successful request.
\begin{equation}
e_{\mathrm{req}}=\frac{\sysenergy}{N_{\mathrm{success}}}
\quad [\mathrm{J/request}],
\label{eq:method-energy-per-request}
\end{equation}
where $\sysenergy$ is the combined energy of the three serving GPUs over the measurement window and $N_{\mathrm{success}}$ is the number of successfully completed requests.

\begin{figure}[b]
  \centering
  % Keep paragraph spacing out of the image-to-caption gap.
  % \setlength{\parskip}{0pt}
  \includegraphics[width=\columnwidth]{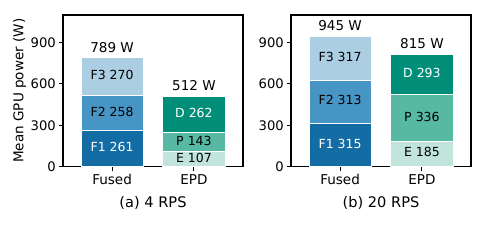}\par\nointerlineskip
  \caption{Mean GPU power for Qwen3-VL on VisionArena at (a) 4~RPS and (b) 20~RPS. F1--F3 denote fused replicas. E, P, and D denote dedicated stage GPUs. Lower is better.}
  \label{fig:part-a-power}
\end{figure}

\begin{figure*}[t]
  \centering
  % Keep paragraph spacing out of the image-to-caption gap.
  \setlength{\parskip}{0pt}
  \includegraphics[width=\textwidth]{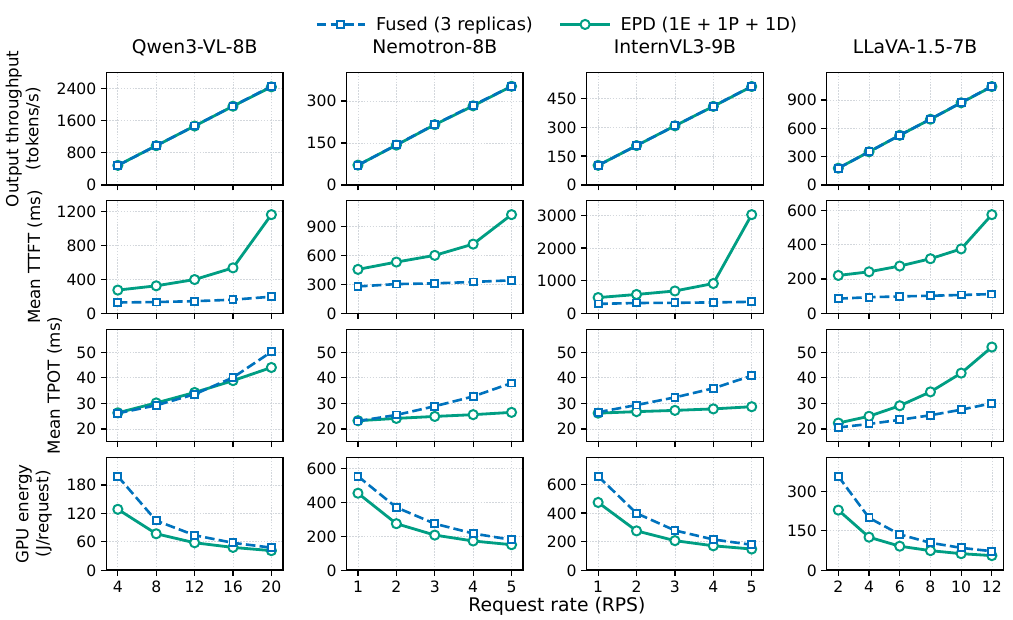}\par\nointerlineskip
  \caption{Three fused replicas versus 1E+1P+1D on VisionArena-Chat. Columns show models. Higher is better for output throughput; lower is better for mean TTFT, mean TPOT, and GPU energy. }
  \label{fig:part-a-overview}
\end{figure*}

\section{EPD Disaggregation: Energy and Performance}
\label{sec:disaggregation}
    We compare three fused replicas with a fully disaggregated EPD pipeline under the same three-GPU budget. Figure~\ref{fig:part-a-overview} compares the four models on VisionArena-Chat. Across 21 paired operating points, disaggregation reduces GPU energy per successful request but increases mean TTFT. Its effect on TPOT varies across models and request rates, making deployment selection a joint energy--latency decision.
    
    \subsection{Disaggregation Reduces GPU Energy per Request}
    \label{sec:part-a-energy}
    Across the evaluated loads, disaggregation saves GPU energy at almost the same output-token throughput as fused serving, as shown in Fig.~\ref{fig:part-a-overview}. It reduces GPU energy per request by 13.3--35.2\% for Qwen3-VL, 16.2--26.0\% for Nemotron, 16.6--30.9\% for InternVL3, and 23.1--37.0\% for LLaVA. Paired output throughput differs by less than 0.7\%.
    
    The lower energy consumption accompanies lower combined average GPU power over nearly identical benchmark durations. Figure~\ref{fig:part-a-power}(a) shows how stage separation changes the power distribution. At 4~RPS, the three Qwen3-VL fused replicas draw 260.9, 258.2, and 270.1~W, whereas the dedicated E, P, and D GPUs draw 107.0, 142.6, and 261.9~W. Total average power consequently falls from 789.1 to 511.5~W. In fused serving, requests are distributed across three complete EPD replicas that process their assigned requests in parallel. Each GPU handles encoding and prefill as well as sustained autoregressive decoding, keeping all three GPUs nearly fully utilized at this load. Disaggregation concentrates generation on D, while E and P process incoming inputs without sustaining autoregressive decoding. At this load, the sampled GPU busy fractions are 6.7\% for E and 21.3\% for P, compared with approximately 100\% for D and each fused replica. The longer inactive intervals at E and P help explain their lower average power and the corresponding reduction in total GPU energy.

The energy advantage narrows at higher loads as the dedicated stages become busier. As illustrated in Fig.~\ref{fig:part-a-power}, from 4 to 20~RPS, Qwen3-VL's E and P average power rises from 107.0 and 142.6~W to 185.3 and 336.2~W, respectively, while D rises from 261.9 to 293.5~W. The E and P busy fractions also increase to 34.7\% and 94.6\%. Aggregate disaggregated power therefore increases from 511.5 to 814.9~W, while fused power increases from 789.1 to 945.4~W. This faster increase in disaggregated power narrows the energy saving from 35.2\% to 13.3\%. The other models also show smaller savings toward their highest evaluated loads, although the full curves are not uniformly monotonic. Higher stage activity reduces the power advantage. Disaggregation remains more energy-efficient at all retained points.

\subsection{Disaggregation Increases Mean TTFT}
\label{sec:part-a-ttft}
Disaggregation increases mean TTFT, and the gap generally grows with request rate. As shown in Fig.~\ref{fig:part-a-overview}, for Qwen3-VL, mean TTFT increases from 0.128 to 0.274~s at 4~RPS and from 0.196 to 1.159~s at 20~RPS. The same pattern appears in the other models. At their highest loads, mean TTFT for Nemotron increases from 0.343 to 1.024~s at 5~RPS, InternVL3 from 0.355 to 3.024~s at 5~RPS, and LLaVA from 0.112 to 0.575~s at 12~RPS.

Separation lengthens the request's execution path across instances. Before generation begins, visual representations must pass from E to P, followed by KV state from P to D, adding transfers, coordination, and waits to the first-token path. Resource allocation also changes the queueing behavior. Three fused replicas distribute arrivals across three complete processing paths, whereas each dedicated stage in 1E+1P+1D receives the full request stream. As load increases, the stage with insufficient processing capacity accumulates requests, and spare capacity on another stage's GPU cannot directly clear that queue. The additional handoffs and growing stage queues together explain why the mean TTFT penalty grows at higher request rates.

\subsection{TPOT Reflects E/P Interference and Decode Resource Pressure}
\label{sec:part-a-tpot}
Disaggregation generally improves TPOT for workloads with high E/P demand, as shown in Fig.~\ref{fig:part-a-overview}. For Nemotron at 5~RPS, mean TPOT falls from 37.95 to 26.49~ms, a 30.2\% reduction. For InternVL3 at 5~RPS, it has a 29.7\% reduction. Qwen3-VL shows smaller differences at lower loads, with TPOT remaining close to fused serving at 4--12~RPS. At higher request rates, disaggregation reduces its TPOT by 3.0\% at 16~RPS and 12.2\% at 20~RPS.

Separating E and P from D removes their direct competition for the same GPU's compute resources. In fused serving, encoding and prefill chunks share GPU resources with ongoing decoding. Chunked prefill limits the amount of prefill work scheduled at a time, but this work still competes with generation. After disaggregation, E and P execute on other GPUs, allowing D to process output tokens without sharing its GPU with visual encoding and prompt processing. This isolation provides greater benefit when E/P work imposes substantial interference, consistent with the larger TPOT improvements observed for Nemotron and InternVL3.

LLaVA is an exception. Disaggregation increases its TPOT at every evaluated request rate. Its relatively light encoding limits the interference that can be removed, while its larger per-token KV footprint and the concentration of all decode requests on one D GPU increase resource pressure. Without preemptions, its TPOT penalty already grows from 8.4\% at 2~RPS to 51.9\% at 10~RPS. At 12~RPS, sampled D-side KV-cache utilization reaches 100\%, with 18 preemptions and a 73.2\% TPOT increase. The TPOT outcome therefore reflects a balance between E/P interference with decoding in fused serving and competition among decode requests after separation. Isolation improves TPOT when removing the former outweighs increasing the latter.

\subsection{Choosing Between Fused and Disaggregated Serving}
\label{sec:part-a-choice}

Prefer disaggregation when energy efficiency or steady token generation is a priority and additional first-token delay is tolerable. This recommendation is supported by the 13.3--37.0\% reduction in GPU energy per successful request across our evaluated operating points, while output-token throughput differs by less than 0.7\% from fused serving, as shown in Fig.~\ref{fig:part-a-overview}. Workloads with substantial E/P demand are particularly promising because a dedicated D GPU can reduce interference during generation, as illustrated by the TPOT improvements for Nemotron and InternVL3. Disaggregation is therefore attractive when its additional TTFT remains within the application's SLO. Potential use cases include noninteractive agentic workflows with relaxed first-response deadlines, or large-image analysis whose response-time budget accommodates substantial visual processing.

Prefer fused serving when latency is the priority and the SLO leaves little room for additional handoffs or queueing. For example, Qwen3-VL at 20~RPS sees mean TTFT increase from 0.196 to 1.159~s after disaggregation, illustrating the additional first-response delay that an application must accommodate. When E/P work is light enough to coexist with decoding on a fused GPU, separation offers less interference reduction. Concentrating requests at a dedicated stage can also introduce queueing, decode contention, or KV-cache pressure~\cite{jiang2026}. The 73.2\% TPOT increase for LLaVA at 12~RPS illustrates the potential generation-time cost. Under the tested three-GPU budget, fused serving distributes requests across three complete replicas. Each request executes E, P, and D within its assigned replica, avoiding the additional inter-instance transfers and cross-stage coordination introduced by separation.

Once an application selects EPD disaggregation, the next question is how to apply stage-wise DVFS to further reduce whole-pipeline GPU energy. Although the stages can be clocked independently, their queues and execution times remain coupled. The next section therefore examines how the energy--performance trade-off changes with the stage being downclocked, request rate, workload content, and model, identifying opportunities for further savings within the latency budget specified by the application.

\begin{figure*}[!t]
  \centering
  \includegraphics[width=\textwidth]{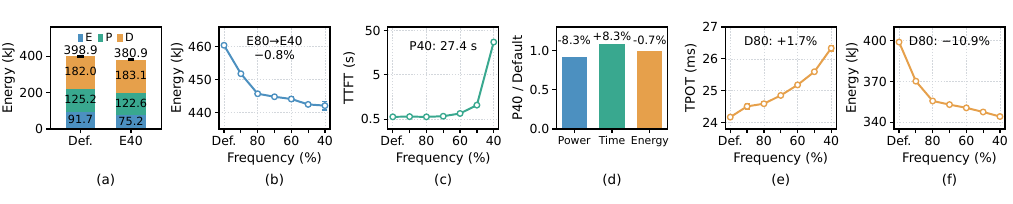}% \par\nointerlineskip
  \caption{Stage-specific DVFS effects on VisionArena-Chat in the 1E+1P+1D pipeline. (a) Energy of E, P, and D at Default and E40 (Nemotron, 2~RPS). (b) E-only energy (Qwen3-VL, 10~RPS). (c,d) P-only TTFT and P40/Default power, benchmark duration, and energy ratios (Qwen3-VL, 15~RPS). (e,f) D-only TPOT and energy (Nemotron, 2~RPS). Lower is better for all metrics. (c) uses a logarithmic axis.}
  \label{fig:dvfs-stage-effects}
\end{figure*}

\section{Stage-Wise DVFS in EPD Serving}
\label{sec:dvfs}
\begin{figure*}[!t]
  \centering
  \setlength{\parskip}{0pt}
  \begin{minipage}[t]{0.49\textwidth}
    \vspace{0pt}
    \centering
    \includegraphics[width=\linewidth]{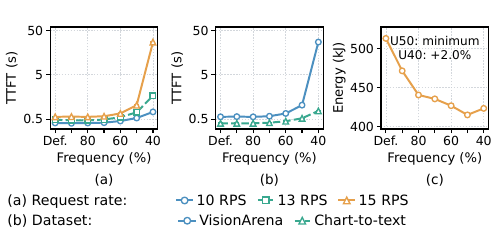}\par\nointerlineskip
    \caption{Workload effects for Qwen3-VL. (a) P-only mean TTFT across VisionArena-Chat request rates. (b) P-only mean TTFT across datasets at 15~RPS. (c) Uniform energy on Chart-to-text at 20~RPS. Lower is better.}
    \label{fig:dvfs-workload-effects}
  \end{minipage}\hfill%
  \begin{minipage}[t]{0.49\textwidth}
    \vspace{0pt}
    \centering
    \includegraphics[width=\linewidth]{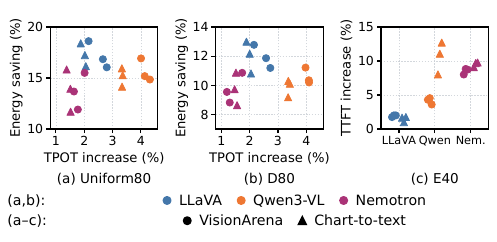}\par\nointerlineskip
    \caption{Cross-model DVFS effects across 18 workloads. (a) Uniform80 and (b) D80 energy savings versus mean TPOT increases. (c) E40 mean TTFT increases. Higher is better for energy saving; lower is better for TTFT and TPOT increases.}
    \label{fig:dvfs-model-effects}
  \end{minipage}
\end{figure*}

We fix the 1E+1P+1D allocation and examine how stage-wise downclocking changes whole-pipeline energy and latency. The experiments cover Qwen3-VL~\cite{qwen}, Nemotron~\cite{nemotron}, and LLaVA~\cite{llava} on VisionArena-Chat~\cite{visionarena} and Chart-to-text~\cite{charttotext,onevisiondata}. Three request rates per model--dataset pair yield 18 workload combinations. Each frequency setting is compared with the default-frequency disaggregated baseline for the same workload. E80 denotes an 80\% frequency target for E alone. Uniform80 applies that target to all three stages. Latencies below are means unless stated otherwise. Figures~\ref{fig:dvfs-stage-effects}, \ref{fig:dvfs-workload-effects}, and~\ref{fig:dvfs-model-effects} present representative evidence for the insights. Complete results for all 18 workloads are provided in App.~\ref{app:all-dvfs-results}.

\subsection{Stage-Specific Responses to Downclocking}
\label{sec:dvfs-rq1}
Stage responses to downclocking differ. E offers limited whole-pipeline savings, deep P downclocking can trigger queueing that offsets lower power, and moderate D downclocking provides larger savings with small latency costs.
%
% \insight{1.1}{Lowering E frequency provides limited whole-pipeline energy savings, with diminishing returns from further reductions.}
%
\begin{takeaway}
\label{takeaway1} 
\textit{Lowering E frequency provides limited whole-pipeline energy savings, with diminishing returns from further reductions.}
\end{takeaway}
E's local energy saving has limited whole-pipeline impact because P and D consume most of its energy. For example, with Nemotron + VisionArena at 2 RPS (Fig.~\ref{fig:dvfs-stage-effects}(a)), E40 reduces E's energy from 91.7 to 75.2\,kJ, an 18.0\% reduction, but lowers total GPU energy by only 4.5\%. At default, E, P, and D account for 23\%, 31\%, and 46\% of total energy. Thus, other GPUs' consumption dilutes the saving at E.

Further E downclocking produces little additional whole-pipeline saving. For Qwen3-VL + VisionArena at 10 RPS (Fig.~\ref{fig:dvfs-stage-effects}(b)), reducing E from E80 to E40 lowers total energy from 445.7 to 442.1\,kJ, only another 0.8\%. Telemetry confirms that E's mean clock falls from 1,989 to 1,005\,MHz, whereas its average board power falls only from 110.1 to 106.1\,W. P and D retain their default frequency settings. Slower encoding can lengthen active computation, limiting the power benefit of a lower clock. Combined with P and D's larger energy shares, this leaves little additional whole-pipeline saving from further E downclocking.
%
% \insight{1.2}{Deep P downclocking can trigger a TTFT cliff and prolong the benchmark, offsetting power savings.}
\begin{takeaway}
\label{takeaway2}
\textit{Deep P downclocking can trigger a TTFT cliff and extend runtime, offsetting power savings.}
\end{takeaway}
%
% P becomes highly sensitive to downclocking when its processing capacity approaches the incoming workload. For Qwen3-VL + VisionArena at 15 RPS (Figure~\ref{fig:dvfs-stage-effects}(c)), P40 raises TTFT from 0.565 to 27.42\,s. Supporting P logs record peak waiting-request counts of 17 at default and 775 at P40. Reducing P's capacity allows requests to accumulate, producing a sharp increase in first-token latency.

P becomes highly sensitive to downclocking when its processing capacity approaches the incoming workload. For Qwen3-VL + VisionArena at 15~RPS (Fig.~\ref{fig:dvfs-stage-effects}(c)), P70 is the lowest tested P-only setting that satisfies both latency SLOs across all three repetitions. At P60, P99 TTFT reaches 2.13--2.17\,s, exceeding the 2\,s budget. Further downclocking to P40 raises mean TTFT from 0.565 to 27.42\,s. Supporting P logs record peak waiting-request counts of 17 at default and 775 at P40. Reducing P's capacity allows requests to accumulate, producing a sharp increase in first-token latency.

A longer benchmark can offset the reduction in GPU power. In the same experiment (Fig.~\ref{fig:dvfs-stage-effects}(d)), average aggregate GPU power falls from 766.4 to 702.4\,W, but benchmark duration grows from 672 to 728\,s. Total energy therefore falls only from 515.0 to 511.3\,kJ, a 0.7\% saving. The relationship is
\begin{equation}
\sysenergy=\overline{P}_{\mathrm{GPU}}\,T_{\mathrm{benchmark}},
\label{eq:dvfs-power-time}
\end{equation}
where $\overline{P}_{\mathrm{GPU}}$ is the three GPUs' combined average power. All three GPUs consume energy while queued requests drain, so the longer benchmark largely offsets lower power.

\begin{takeaway}
\label{takeaway3}
\textit{Moderate D downclocking provides substantial energy savings with small TPOT penalties.}
\end{takeaway}
Moderate D downclocking reduces energy with little change in token-generation speed. For Nemotron + VisionArena at 2 RPS (Fig.~\ref{fig:dvfs-stage-effects}(e,f)), D80 raises TPOT by only 1.7\% while reducing total energy by 10.9\%. D40 increases the saving to 13.7\%, but its TPOT penalty reaches 8.9\%.

This behavior is consistent with the memory-intensive nature of decoding identified in prior work. Splitwise~\cite{splitwise} shows that token generation can tolerate lower power provision, while DynamoLLM~\cite{dynamollm} shows that decode-heavy workloads can benefit from lower GPU frequencies with small performance losses. When memory bandwidth limits decoding, moderate core downclocking need not proportionally slow token generation. Our EPD results exhibit the same favorable pattern when D is isolated from both E and P.

\subsection{Effects of Request Rate and Workload Content}
\label{sec:dvfs-rq2}
Both request rate and per-request work determine the cost of downclocking. Higher loads make queueing more sensitive to frequency reductions. Once benchmark duration increases, the lowest frequency need not minimize energy.

\begin{takeaway}
\label{takeaway4}
\textit{Higher request rates make the same P downclocking cause much larger TTFT increases.}
\end{takeaway}
The same P frequency target produces sharply different TTFT penalties as request rate increases. For Qwen3-VL + VisionArena (Fig.~\ref{fig:dvfs-workload-effects}(a)), P40 increases TTFT by 77\% at 10 RPS, 249\% at 13 RPS, and 4,756\% at 15 RPS, reaching 0.73, 1.69, and 27.42\,s, respectively. Higher arrival rates leave P with less spare capacity. The same frequency reduction can then push it into sustained request accumulation, amplifying the associated TTFT penalty.

\begin{takeaway}
\label{takeaway5}
\textit{Different datasets produce different TTFT penalties at the same request rate and P frequency.}
\end{takeaway}
Matching request rates does not produce matching workloads at P. For Qwen3-VL at 15 RPS (Fig.~\ref{fig:dvfs-workload-effects}(b)), P40 raises TTFT from 0.565 to 27.42\,s on VisionArena, but from 0.402 to 0.767\,s on Chart-to-text, increases of 4,756\% and 90\%, respectively. Different image workloads create different visual-processing demands and visual-token counts. Together with text-context lengths, these change the amount of prefill work per request, so the same P frequency and RPS can produce very different queueing delays.

\begin{samepage}
\begin{takeaway}
\label{takeaway6}
\textit{Energy is minimized by balancing power and benchmark duration. Further downclocking can increase total energy.}
\end{takeaway}
\end{samepage}
The lowest GPU frequency is not necessarily the lowest-energy operating point. For Qwen3-VL + Chart-to-text at 20 RPS (Fig.~\ref{fig:dvfs-workload-effects}(c)), total energy reaches its measured minimum at Uniform50, at 415.1\,kJ. Uniform40 lowers average aggregate power further, from 632.7 to 588.2\,W, but extends the benchmark from 656 to 720\,s. Energy consequently rises to 423.5\,kJ, 2.0\% above Uniform50, while TTFT grows from 1.00 to 31.69\,s. The increase in benchmark duration outweighs the corresponding power reduction.

Moderate uniform downclocking provides savings with small mean-latency changes across the tested workloads. Across all 18 combinations (Fig.~\ref{fig:dvfs-model-effects}(a)), Uniform80 reduces total energy by 11.7--18.6\%, with TTFT changes of $-1.1$ to $+2.7$\% and TPOT increases of 1.4--4.3\%.

\subsection{Cross-Model Generality of Stage-Wise Trends}
\label{sec:dvfs-rq3}
\begin{takeaway}
\label{takeaway7}
\textit{Qualitative stage-wise trends recur across models, but frequency targets and sustainable request rates require calibration.}
\end{takeaway}
Moderate D downclocking consistently offers useful energy savings across the evaluated models and datasets. Across all 18 combinations (Fig.~\ref{fig:dvfs-model-effects}(b)), D80 reduces total energy by 8.6--13.0\%, with TPOT increases of only 1.2--4.1\%.

Encoder sensitivity differs across models, which use different visual encoders and image-processing designs. Under E40 (Fig.~\ref{fig:dvfs-model-effects}(c)), TTFT increases on VisionArena are 1.7--2.0\% for LLaVA, 3.6--4.5\% for Qwen3-VL, and 8.0--8.8\% for Nemotron. On Chart-to-text, Qwen3-VL's increase is 8.0--12.7\%. Different encoders and input-dependent visual workloads therefore lead to different end-to-end latency responses to the same E downclocking target.

What generalizes is the recurring qualitative pattern. Moderate D downclocking can save substantial energy, while E's latency sensitivity depends on the model and dataset. The same frequency reduction does not produce the same performance penalty or energy saving. Frequency targets and sustainable request rates therefore need to be calibrated for each model and its target request workload.

\section{Practical Frequency Selection}
\label{sec_recommend}

The preceding section shows that the energy-latency effects of downclocking depend on the stage and workload. In this section, we seek to identify an energy-efficient configuration under P99 TTFT and mean TPOT SLOs using only a few profiling measurements. For a given workload, we propose a simple yet effective exploration strategy to select a configuration for the fixed 1E+1P+1D pipeline from the 25 configurations defined in \S\ref{sec:method-framework}.

%that prioritizes Uniform scaling and selectively considers stage-only alternatives
%This motivates a practical question: how can we select an energy-efficient frequency configuration under given latency requirements without exhaustively profiling all candidate configurations? In this section, we translate our characterization into a

\subsection{Recommended Exploration}

The exploration evaluates at most $K$ distinct configurations, where $K$ is a user-specified budget. Default and infeasible probes both count toward $K$. The procedure maintains the lowest-energy SLO-feasible configuration found after each probe, allowing a budget-aware \textit{anytime} recommendation. The exploration addresses three decisions, \textit{where to start}, \textit{how to adjust frequency}, and \textit{when to switch control modes}. 

%A smaller user budget $K$ needs a quick recommendation, while a larger $K$ allows further exploration and potentially greater energy savings. 

\textit{(1) Start with moderate Uniform downclocking}. We first measure Default as the reference, followed by U80. Across the tested 18 workloads, U80 reduces whole-pipeline energy with modest changes in mean latency (Takeaway~\ref{takeaway6}), making it a promising initial downclocking probe.

\textit{(2) Adjust frequency using measured feedback.} Since lower frequencies do not always reduce energy (Takeaway~\ref{takeaway6}), we use a bisection-like heuristic search over frequency levels from 40\% to 100\%. When a probe satisfies both SLOs and improves energy over the current feasible reference, the search moves toward lower frequencies. Otherwise, it moves toward higher frequencies. A feasible probe also establishes a new reference when the previous one is infeasible. For example, if U80 is feasible and reduces energy, the next probe is U60. If U60 is infeasible, the search backs to U70.

\textit{(3) Consider stage-only control selectively.} If U80 is infeasible under a TPOT budget no greater than Default's mean TPOT, we directly explore E-only configurations starting from E60, avoiding further downclocking of P and D. Otherwise, we continue the Uniform search. If any Uniform probe violates an SLO, we explore moderate D-only configurations, followed by E-only if the budget permits. This ordering reflects D's favorable energy-latency trade-off (Takeaway~\ref{takeaway3}) and E's more limited total energy savings (Takeaway~\ref{takeaway1}). If Uniform exploration encounters no SLO violation, we terminate without stage-only exploration. Given P’s queueing risk (Takeaway~\ref{takeaway2}), P80 is reserved for cases where the best feasible configuration remains Default or no feasible configuration has been found. 

The procedure retains the lowest-energy feasible configuration measured, or no recommendation if none is feasible. Detailed probing and stopping rules are provided in App.~\ref{app_recommend}.

\subsection{Effectiveness of Budgeted Exploration}

We evaluate the effectiveness of the strategy across 18 workloads studied in \S\ref{sec:dvfs} under two families of SLOs, (i) the original absolute budgets of P99 TTFT $\leq$ 2s and mean TPOT $\leq$ 50ms, and (ii) workload-specific budgets obtained by adding margins of 0, 5\% to Default's P99 TTFT and 10\%, 20\%, 50\%, and 0, 2\%, 5\%, 10\%, 20\% to its mean TPOT, using averages across Default repetitions. The latter consists of 288 positive-margin scenarios, where both margins are positive, and 162 zero-margin scenarios, where at least one margin is zero. A configuration is feasible only if all valid repetitions satisfy both SLOs. We compare the configuration selected within budget $K$ against the measured oracle, defined as the lowest-energy feasible configuration among all 25 candidates. Table~\ref{tab:budgeted_exploration} summarizes the results.

\begin{table}[t]
\centering
\caption{Budgeted exploration against the measured oracle. Under absolute (abs.) and positive-margin (pos.) SLOs, we report median/maximum energy gaps to the oracle. Lower is better. Under zero-margin (Zero) SLOs, we report the number of feasible scenarios found out of 107. Higher is better.}
\label{tab:budgeted_exploration}
\small
\begin{tabular}{@{}lcccc@{}}
\toprule
Method & $K$ & Abs. gap (\%) & Pos. gap (\%) & Zero (\#) \\
\midrule
Ours & 2 & 5.67 / 6.93 & 1.49 / 13.49 & 16 \\
& 3 & 2.19 / 3.77 & 0.00 / 7.36 & 43 \\
& 4 & 0.00 / 2.15 & 0.00 / 0.00 & 56 \\
& 5 & 0.00 / 0.00 & 0.00 / 0.00 & 56 \\
& 6 & 0.00 / 0.00 & 0.00 / 0.00& 76 \\
\midrule
U-only & 4 & 0.00 / 2.15 & 0.00 / 7.36 & 23 \\
& 6 & 0.00 / 0.00 & 0.00 / 7.36 & 23 \\
\bottomrule
\end{tabular}
\end{table}

Under the absolute SLOs, 17 of the 18 workloads admit feasible configurations. With only $K=2$, our strategy finds feasible configurations for all 17, achieving 15.7\% average energy savings over Default with a maximum oracle gap of 6.9\%. The gap further decreases with additional probes, reaching zero for all 17 workloads at $K=5$. For the 288 positive-margin scenarios, $K=2$ already finds feasible recommendations in every case with a median oracle gap of 1.5\%, and $K=4$ matches the oracle in all scenarios. 

Zero-margin SLOs are more challenging. Only 107 of the 162 scenarios admit feasible configurations, partly because thresholds based on Default averages may not be satisfied by every repetition. At $K=6$, our strategy identifies feasible configurations in 76 of these scenarios compared with 23 for Uniform-only exploration. The remaining misses illustrate the non-monotonicity of SLO feasibility across frequencies, which can cause the heuristic to overlook feasible candidates. Overall, the strategy recovers the measured oracle with four to five probes under the absolute and positive-margin SLOs, compared with exhaustive exploration of 25 configurations.

\section{Conclusion}
\label{sec:conclusion}
This paper studies whole-service GPU energy and performance trade-offs in multimodal online inference at two levels, deployment and stage frequency. Under a fixed three-GPU budget, disaggregation reduces energy across the tested operating points, but additional handoffs, state transfers, and queueing increase mean TTFT. Token-generation performance depends on the balance between stage-isolation benefits and resource pressure at the dedicated decode instance. Downclocking must likewise be evaluated across the complete pipeline. E's contribution to energy savings is limited by its share of total energy, deep P downclocking can offset power savings through queueing and longer execution, and moderate D downclocking generally saves energy at a small performance cost. Models, workloads, and latency budgets jointly determine the available scope for downclocking, so the lowest frequency need not minimize energy. These findings guide budgeted exploration, which matches the lowest feasible energy among 25 measured candidates with four to five configuration probes under the evaluated absolute and positive-margin SLOs.

% Existing figures can be included from figures/ using \includegraphics.
% Reuse the project bibliography for this writing entry point.
\bibliography{example_paper}
\bibliographystyle{mlsys2025}

\clearpage

\appendix

\section{AI Usage Disclosure}
\label{app:ai-usage}
We used OpenAI's ChatGPT for grammar correction and writing assistance. All data collection, experiments, and validation were performed by the authors. The authors take full responsibility for the content of this paper.

\section{RQ2: Additional DVFS Results}
\label{app:all-dvfs-results}

The main text selects representative operating points to explain the effects of stage-wise downclocking on whole-pipeline energy and performance. This appendix provides results for all 18 workload combinations from three models, two datasets, and three request rates. Each workload includes Default and complete frequency sweeps for single-stage E, P, and D downclocking and uniform downclocking. The figures show, in order, total energy across the three GPUs, output-token throughput, mean TTFT, and mean TPOT. Each point is the mean across benchmark repetitions. Each configuration is compared with the default-frequency result for the same model, dataset, and request rate.

\subsection{Qwen3-VL-8B}

\begin{figure*}[tp]
    \centering
    \includegraphics[width=\textwidth]{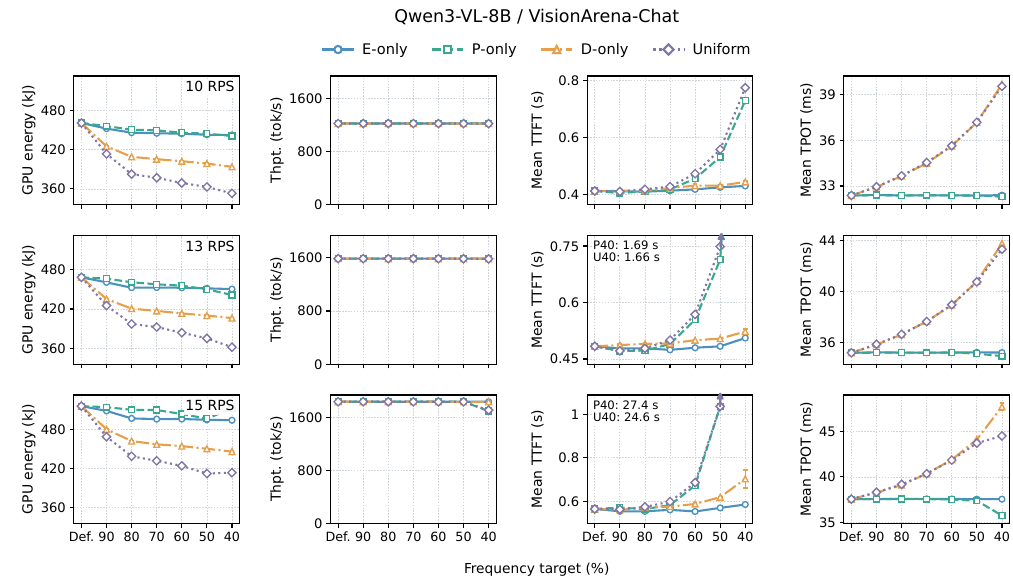}
    \vspace{-8pt}
    \caption{Complete DVFS results for Qwen3-VL-8B on VisionArena-Chat. We report three-GPU energy, output-token throughput, mean TTFT, and mean TPOT. Points show configuration means.}
    \label{fig:appendix-dvfs-qwen-visionarena}
\end{figure*}

\begin{figure*}[tp]
    \centering
    \includegraphics[width=\textwidth]{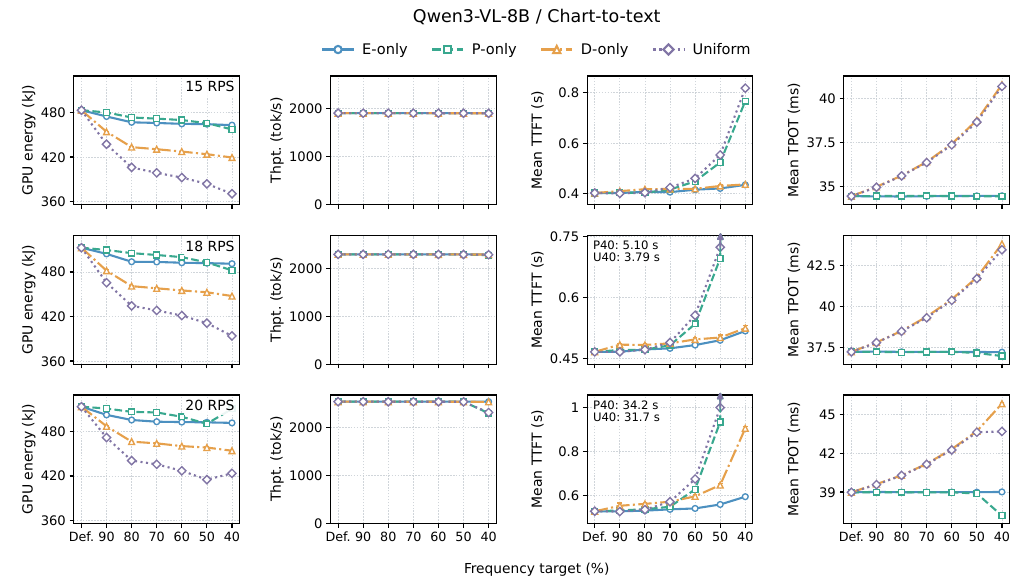}
    \vspace{-8pt}
    \caption{Complete DVFS results for Qwen3-VL-8B on Chart-to-text. We report three-GPU energy, output-token throughput, mean TTFT, and mean TPOT. Points show configuration means.}
    \label{fig:appendix-dvfs-qwen-charttotext}
\end{figure*}

The complete results for Qwen3-VL (Figures~\ref{fig:appendix-dvfs-qwen-visionarena} and~\ref{fig:appendix-dvfs-qwen-charttotext}) are consistent with the main-text observation that E downclocking provides limited whole-pipeline energy savings. Across the six workloads from the two datasets, E80 reduces total energy across the three GPUs by 3.2\%--3.7\%, while further downclocking to E40 yields reductions of only 3.8\%--4.3\%. Although the encoder's target frequency continues to decrease, the additional reduction in whole-pipeline energy is small, indicating that the marginal energy savings from E downclocking diminish relatively early in these workloads.

Deep P downclocking and uniform downclocking show stronger load dependence. On VisionArena-Chat, as the request rate increases from 10~RPS to 15~RPS, mean TTFT under P40 rises from 0.73 to 27.42~s. The uniform frequency with the lowest energy also changes with load. Four workloads at lower or moderate loads reach their lowest measured total energy at Uniform40, whereas the workload at the highest load from each dataset reaches its minimum at Uniform50. The full sweeps thus supplement the main-text observation that deeper downclocking at higher loads can cause substantial latency increases and does not necessarily further reduce total GPU energy consumption.

% \noindent\begin{minipage}{\textwidth}
%   \centering
%   \setlength{\parskip}{0pt}
%   \includegraphics[width=\textwidth]{figures/appendix_dvfs_qwen_visionarena.pdf}\par\nointerlineskip
%   \vspace{2pt}
%   \includegraphics[width=\textwidth]{figures/appendix_dvfs_qwen_charttotext.pdf}\par\nointerlineskip
%   \makeatletter
%   \def\@captype{figure}
%   \makeatother
%   \vspace{-6pt}
%   \caption{Complete DVFS results for Qwen3-VL-8B on VisionArena-Chat (top) and Chart-to-text (bottom). Rows show request rates; columns show three-GPU energy, output-token throughput, mean TTFT, and mean TPOT. Points show configuration means.}
%   \label{fig:appendix-dvfs-qwen}
% \end{minipage}

\subsection{Nemotron-8B}

\begin{figure*}[tp]
    \centering
    \includegraphics[width=\textwidth]{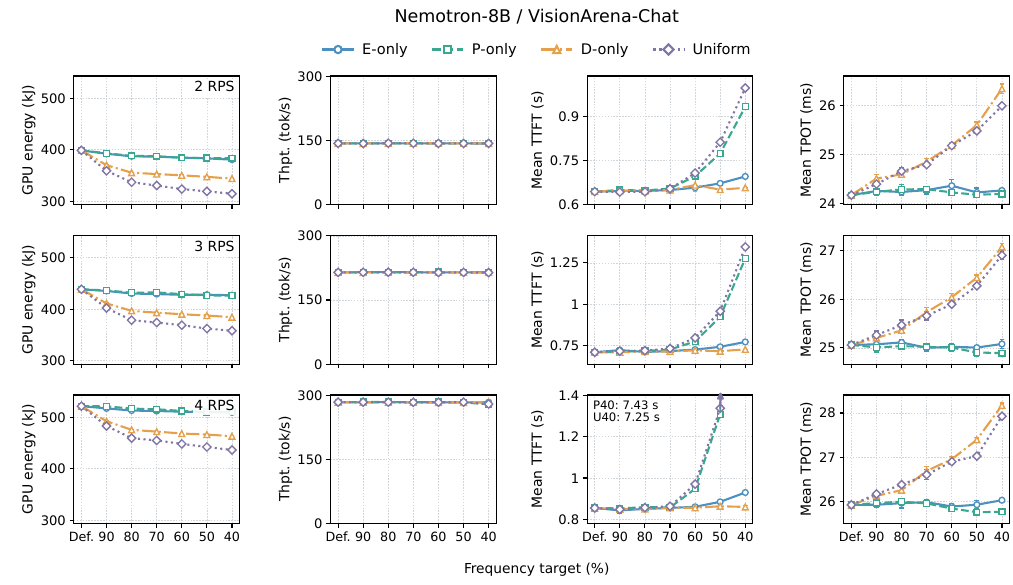}
    \vspace{-8pt}
    \caption{Complete DVFS results for Nemotron-8B on VisionArena-Chat. We report three-GPU energy, output-token throughput, mean TTFT, and mean TPOT. Points show configuration means.}
    \label{fig:appendix-dvfs-nemotron-visionarena}
\end{figure*}

\begin{figure*}[tp]
    \centering
    \includegraphics[width=\textwidth]{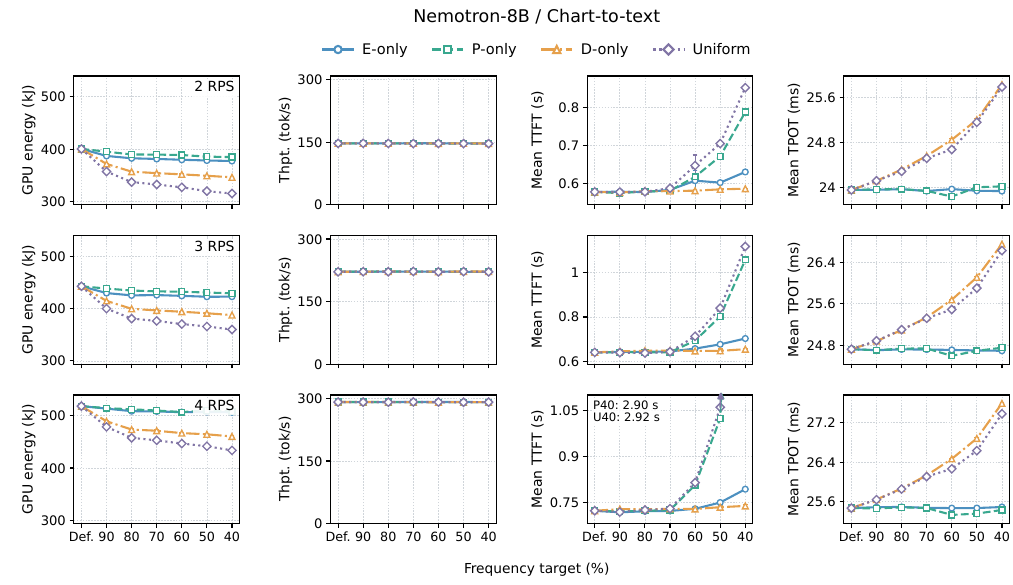}
    \vspace{-8pt}
    \caption{Complete DVFS results for Nemotron-8B on Chart-to-text. We report three-GPU energy, output-token throughput, mean TTFT, and mean TPOT. Points show configuration means.}
    \label{fig:appendix-dvfs-nemotron-charttotext}
\end{figure*}

The complete results for Nemotron (Figures~\ref{fig:appendix-dvfs-nemotron-visionarena} and~\ref{fig:appendix-dvfs-nemotron-charttotext}) further show how downclocking different stages affects different latency metrics. Across the six workloads, E40 increases mean TTFT by 8.0\%--9.8\%, while the magnitude of the change in mean TPOT remains below 0.5\%. By comparison, D80 reduces total energy by 8.6\%--10.9\%, with a mean TPOT increase of only 1.2\%--1.7\%. These results are consistent with the stage-sensitivity analysis in the main text. The performance cost of E downclocking is more apparent in the initial response, while moderate D downclocking achieves whole-pipeline energy savings with only a small cost in subsequent token-generation latency.

The latency cost of deep P downclocking depends on both request rate and request content. On VisionArena-Chat, mean TTFT under P40 increases from 0.93~s at 2~RPS to 7.43~s at 4~RPS. The same 4~RPS and P40 configuration yields a mean TTFT of 2.90~s on Chart-to-text. Thus, even with the same model and target frequency, different datasets exhibit different magnitudes of latency growth.

% \noindent\begin{minipage}{\textwidth}
%   \centering
%   \setlength{\parskip}{0pt}
%   \includegraphics[width=\textwidth]{figures/appendix_dvfs_nemotron_visionarena.pdf}\par\nointerlineskip
%   \vspace{2pt}
%   \includegraphics[width=\textwidth]{figures/appendix_dvfs_nemotron_charttotext.pdf}\par\nointerlineskip
%   \makeatletter
%   \def\@captype{figure}
%   \makeatother
%   \vspace{-6pt}
%   \caption{Complete DVFS results for Nemotron-8B on VisionArena-Chat (top) and Chart-to-text (bottom).}
%   \label{fig:appendix-dvfs-nemotron}
% \end{minipage}

\subsection{LLaVA-1.5-7B}

\begin{figure*}[tp]
    \centering
    \includegraphics[width=\textwidth]{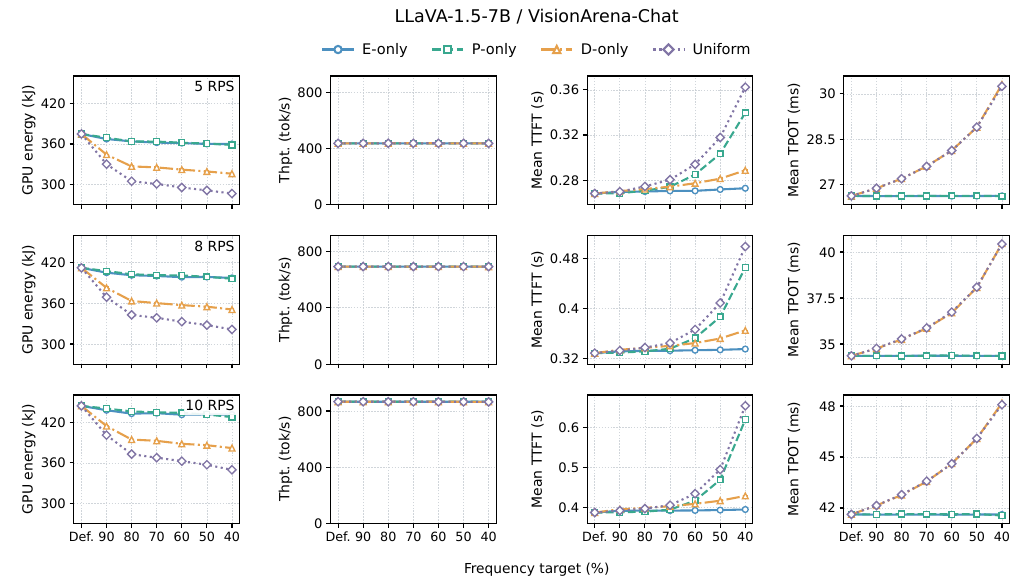}
    \vspace{-8pt}
    \caption{Complete DVFS results for LLaVA-1.5-7B on VisionArena-Chat. We report three-GPU energy, output-token throughput, mean TTFT, and mean TPOT. Points show configuration means.}
    \label{fig:appendix-dvfs-llava-visionarena}
\end{figure*}

\begin{figure*}[tp]
    \centering
    \includegraphics[width=\textwidth]{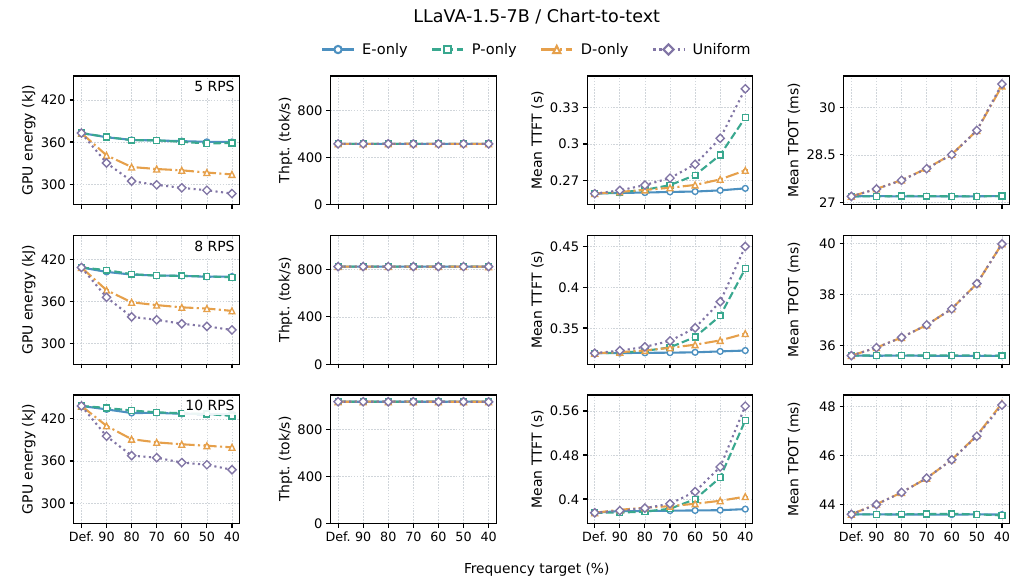}
    \vspace{-8pt}
    \caption{Complete DVFS results for LLaVA-1.5-7B on Chart-to-text. We report three-GPU energy, output-token throughput, mean TTFT, and mean TPOT. Points show configuration means.}
    \label{fig:appendix-dvfs-llava-charttotext}
\end{figure*}

The complete results for LLaVA (Figures~\ref{fig:appendix-dvfs-llava-visionarena} and~\ref{fig:appendix-dvfs-llava-charttotext}) show that its encoding stage has a relatively small latency response to downclocking. Across the six workloads, E40 increases mean TTFT by only about 1.0\%--2.0\%, consistent with the lighter encoding workload characterized in the main text. Although P40 increases mean TTFT by 24.1\%--60.2\%, its absolute value remains at most 0.62~s. Under the tested loads, LLaVA does not exhibit the multi-second TTFT increases observed for Qwen3-VL and Nemotron, showing that the same degree of downclocking has different performance boundaries across models.

Meanwhile, LLaVA also offers energy savings through moderate D downclocking. D80 reduces total energy by 10.8\%--13.0\% across the six workloads, while increasing mean TPOT by only 1.9\%--2.7\%. Further downclocking with Uniform40 yields the lowest measured total energy, but the mean TPOT increase grows to 10.3\%--17.7\%. As discussed in \S\ref{sec:part-a-tpot}, LLaVA's larger per-token KV-cache footprint and the concentration of decode requests on a single D instance create persistent decode resource pressure. Therefore, a lighter encoding workload does not imply that the whole pipeline has the same downclocking headroom.

% \noindent\begin{minipage}{\textwidth}
%   \centering
%   \setlength{\parskip}{0pt}
%   \includegraphics[width=\textwidth]{figures/appendix_dvfs_llava_visionarena.pdf}\par\nointerlineskip
%   \vspace{2pt}
%   \includegraphics[width=\textwidth]{figures/appendix_dvfs_llava_charttotext.pdf}\par\nointerlineskip
%   \makeatletter
%   \def\@captype{figure}
%   \makeatother
%   \vspace{-6pt}
%   \caption{Complete DVFS results for LLaVA-1.5-7B on VisionArena-Chat (top) and Chart-to-text (bottom).}
%   \label{fig:appendix-dvfs-llava}
% \end{minipage}

% \clearpage

\section{Details of Budgeted Frequency Selection}
\label{app_recommend}

This section supplements the implementation overview in \S\ref{sec_recommend} with additional details on measurement, feasibility criteria, and the frequency exploration procedure.

\subsection{Measurement and Feasibility Criteria}
\label{app_criteria}

We consider the same 25 configurations defined in \S\ref{sec:dvfs}, comprising Default and six frequency levels for each of the four control modes, Uniform, E-only, P-only, and D-only. Each configuration measurement consists of all valid benchmark repetitions. For each configuration, we use the mean energy per successful request across valid repetitions to rank configurations. SLO feasibility requires every valid repetition to satisfy both the P99 TTFT and mean TPOT constraints defined in \S\ref{sec:method-slo}. Thus, a configuration is feasible only if the maximum P99 TTFT and maximum mean TPOT across its valid repetitions are both within their respective budgets. The exploration retains the lowest-energy feasible configuration among those measured. If no feasible configuration is found, it returns no recommendation.

For workload-specific SLOs, the TTFT and TPOT budgets are computed by applying the specified percentage margins to the corresponding means across Default repetitions. In particular, a zero-margin budget may exclude Default because the feasibility criterion requires every repetition to satisfy a threshold derived from the repetition mean.

The measured oracle is the lowest-energy feasible configuration among all 25 candidates under the same SLOs and feasibility criteria. The oracle is used exclusively for evaluation and is not available to the exploration procedure.

\subsection{Exploration Procedure}
\label{app_explore}

Algorithm~\ref{alg:budgeted-exploration} summarizes the budgeted exploration procedure. Here, $F(c)$ indicates whether configuration $c$ satisfies both SLOs, and $\overline{P}_0$ denotes the mean TPOT of Default. The procedure maintains the measured configuration set $\mathcal{Q}$ and the lowest-energy feasible configuration $c^\star$ identified so far, with $c^\star=\bot$ indicating that no feasible configuration has been found. For conciseness, the algorithm abstracts configuration measurement and mode-specific exploration into three routines, $\mathrm{Probe}$, $\mathrm{Search1D}$, and $\mathrm{ExploreD}$. We detail their probing and stopping rules below.

\begin{algorithm}[t]
\caption{Budgeted Frequency Exploration}
\label{alg:budgeted-exploration}
\small
\begin{algorithmic}[1]
\REQUIRE Profiling budget $K \geq 1$, P99 TTFT budget $B_T$, mean TPOT budget $B_P$
\ENSURE Lowest-energy feasible measured configuration $c^\star$

\STATE $\mathcal{Q} \gets \emptyset,\quad c^\star \gets \bot$
\STATE $\mathrm{Probe}(\mathrm{Default})$
\IF{$|\mathcal{Q}| < K$}
    \STATE $\mathrm{Probe}(\mathrm{U80})$
\ENDIF
\IF{$|\mathcal{Q}| = K$}
    \STATE \textbf{return} $c^\star$
\ENDIF

\IF{$\neg F(\mathrm{U80}) \land B_P \leq \overline{P}_0$}
    \STATE $\mathrm{Search1D}(\mathrm{E},60)$
\ELSE
    \STATE $v \gets \mathrm{Search1D}(\mathrm{U},80)$
    \IF{$\neg v$}
        \STATE \textbf{return} $c^\star$
    \ENDIF
    \STATE $\mathrm{ExploreD}()$
    \IF{$|\mathcal{Q}| < K$}
        \STATE $\mathrm{Search1D}(\mathrm{E},60)$
    \ENDIF
\ENDIF

\IF{$|\mathcal{Q}| < K \land
    (c^\star = \bot \lor c^\star = \mathrm{Default})$}
    \STATE $\mathrm{Probe}(\mathrm{P80})$
\ENDIF
\STATE \textbf{return} $c^\star$
\end{algorithmic}
\end{algorithm}

\noindent\textbf{Configuration probing.} $\mathrm{Probe}(c)$ measures a configuration only if it has not been measured and $|\mathcal{Q}|<K$. Each new measurement is added to $\mathcal{Q}$, regardless of feasibility. A feasible configuration replaces $c^\star$ only if it has lower energy.

\noindent\textbf{One-dimensional exploration.} $\mathrm{Search1D}(m,s)$ explores mode $m\in\{U,E\}$, representing Uniform and E-only, respectively, starting at frequency $s$. Each search initializes the interval $[40,100]$, with Default as the 100\% reference. A feasible probe becomes the new upper endpoint if it reduces energy relative to the current upper reference, or if that reference is infeasible. Otherwise, the lower endpoint is updated. The next probe is the unmeasured interior frequency closest to the interval midpoint, with ties favoring the higher one. If no interior candidate remains, 40\% is tested only if it remains the lower endpoint and is unmeasured. The search stops when no candidate remains or the budget is exhausted. For Uniform exploration, the routine returns $v$, indicating whether any evaluated Uniform probe violated an SLO. 

\noindent\textbf{Stage-only exploration and termination.}
$\mathrm{ExploreD}()$ first probes D90 if U90 was measured and found infeasible, followed by D80 only if D90 is feasible. Otherwise, it probes D80 first, followed by D90 only if D80 is infeasible. P80 is considered as a final fallback only when no feasible energy-saving configuration has been identified relative to Default, or when no feasible configuration has been found. If Uniform exploration encounters no SLO violation, the procedure terminates without exploring stage-only
configurations.

\section{Limitations}
\label{app:limitations}
This study also has limitations. Our experiments use three L40S GPUs within one server and do not evaluate other hardware scales, stage resource allocations, or partially fused deployments. The frequency experiments cover static single-stage and uniform core downclocking. Combinations that simultaneously assign different frequencies to multiple stages and online adaptive control remain unevaluated.

\end{document}